\documentclass[preprint,review,nofootinbib,12pt]{elsarticle}

\usepackage{natbib,amssymb,amsbsy,fleqn}

 \journal{Optics Communications}

\begin{document}

 \begin{frontmatter}

 \title{Dynamics of physical properties of a single-mode quantized field nonlinearly and non-resonantly interacting with two V-type three-level atoms passing consecutively through a cavity}

 \author{E Faraji $^1$ }


 \author{M K Tavassoly$^{1,2}$\corref{cor1}}

 \ead{mktavassoly@yazd.ac.ir}

 \cortext[cor1]{Corresponding author}

 \address{1- Atomic and Molecular Group, Faculty of Physics, Yazd University, Yazd, Iran\\
2- Photonics Research Group, Engineering Research Center, Yazd University, Yazd, Iran}
\begin{abstract}
In this paper we address the analytical solution of the {\it non-resonant} interaction between two identical V-type three-level atoms passing consecutively through a single-mode cavity field in the presence of {\it intensity-dependent coupling}.  By considering an  identical initial condition for both atoms and an initial coherent field, we find the  analytical solution of the state vector of the entire atom-field system.
Accordingly, we could carefully  investigate the influence of  various parameters in the circumstances of the interacting system on different physical quantities such as the atomic population inversion, atom-field entanglement,  field squeezing, sub-Poissonian statistics and the  Wigner quasi-probability distribution function. In detail, we discuss numerically the influences of the detuning parameters and a particular nonlinearity function on the mentioned quantities and demonstrate that they have substantial effects on the temporal behavior of the above-mentioned nonclassical properties.
\end{abstract}
\end{frontmatter}
  \section{Introduction}\label{sec-intro}
The interaction of a two-level atom with a single-mode quantized electromagnetic field presents one of the most fundamental problems in quantum optics  area \cite{Scully}. The simplest powerful scheme to investigate this interaction is the Jaynes-Cummings model (JCM)\cite{Jaynes}. JCM leads to the prediction of a wide range of experimentally interesting  phenomena. Also, in order to survey the JCM in different concepts, interesting results have been attained according to various generalizations of this model.  We may refer to a few  examples of them as follow: the interaction between two two-level atoms and a single-mode field \cite{liao}, the interaction between N-level atom and (N-1)-mode field \cite{ABDEL}, the interactions of a  multi-level atom and one- or two-mode field \cite{Ting,Janowicz}, multi-photon transitions in the atom-field interaction \cite{baghshahi1,Khalil2,BRusakova,Bashkirov}, intensity-dependent JCM (nonlinear regime) \cite{intensity4,intensity5,intensity6} which in particular we will also deal with in the present paper, different interaction schemes between atoms and electromagnetic field in the presence of a  Kerr medium \cite{kerr2,zait,kerr3}, JCM with electromagnetic field in the presence of converter terms \cite{Khalil,converter2},  JCM in the presence of Stark shift \cite{baghshahi1,Obadaaa,Mao} and finally JCM when the atom-field coupling is position-dependent \cite{motion1,motion2}.\\
 The Fock state, as the most nonclassical state, in general  can be prepared in cavity QED experiments in which passing atoms interact with a high-Q cavity field one-by-one (see page 390 of Ref. \cite{Scully}) . A single-photon Fock state is created in this procedure by an adiabatic successive passage sequence in an optical cavity  \cite{Hagley}.
    In the latter case, the realization of a quantum memory in a cavity QED experiment has been reported which will be useful in quantum information processing operations.
 In this regard, Phoenix and Barnett have presented a scheme to entangle two nonlocal atoms passing successively through a cavity \cite{Phoenix2}, while they never interact directly with each other. It is also emphasized there that, this may be happen even when a measurement occurred on the first  extracted atom, while the second atom is still not entered the cavity. The authors then demonstrate that how such a model can violate the Bell's inequality.
 Considering the above-mentioned literature and in particular the Ref. \cite{Arpita}, motivate us to study how the consecutive passage of two V-type three-level atoms transfers a classical (coherent state) cavity field into a nonclassical one.
In more detail, we consider here a basic model to describe the non-resonant interaction between  two identical V-type three-level atoms passing subsequently through a single-mode coherent field which is considered in the intensity-dependent coupling regime.  Recently, a similar model has been studied, however, with constant coupling and in resonant condition \cite{Arpita}. We attempt to remove these limitations and go further to evaluate the effects of detuning and intensity-dependent coupling on some of  the nonclassical properties of the entire atom-field system.
After obtaining the explicit form of the state vector, at first we pay attention to the variation of the atomic population inversion (as exchange of energy between atom and field)  in which collapse-revival of Rabi oscillations are revealed.
 Then, due to the fact that, (i) recently much attention has been paid to the entanglement phenomenon in various circumstances, (ii),  the JCM and its generalizations are the simplest resource of entangled state, we evaluate the degree of entanglement of the obtained system state.
This notion is known as a key resource in the quantum information theory and so plays a central role in quantum computation, quantum information, quantum cryptography \cite{benenti} and quantum teleportation \cite{bennet,cola}.
Next, as two other important nonclassical criteria, we  examine the squeezing \cite{yuen}, first and second order, as well as  the sub-Poissonian photon statistics of the field \cite{shortmandel} and finally the variation of Wigner quasi-probability distribution function in phase space is presented \cite{Scully}.
We should emphasize that our presentation is quite general and may be considered for arbitrary nonlinearity function. However, to present our numerical results which followed by the related discussion, we have chosen the well-known nonlinearity function $f(n)=\sqrt{n} $ as the intensity-dependent function. This function has been  used frequently in the literature \cite{intensity1,intensity2}.\\
The remainder of  paper is organized as follow: we try to find the explicit form of the state vector of the entire system after passing the second atom from the cavity field in the next section. Then in section 3 we discuss on the atomic population inversion, von Neumann entropy, squeezing effects, Mandel parameter and finally Wigner quasi-probability distribution function. At last, section \ref{Summary} includes a summary and concluding remarks.\\
%
   \section{The model and its solution}\label{Model}
\begin{figure}[htbp]
                                 \centerline{\includegraphics[width=0.95\columnwidth]{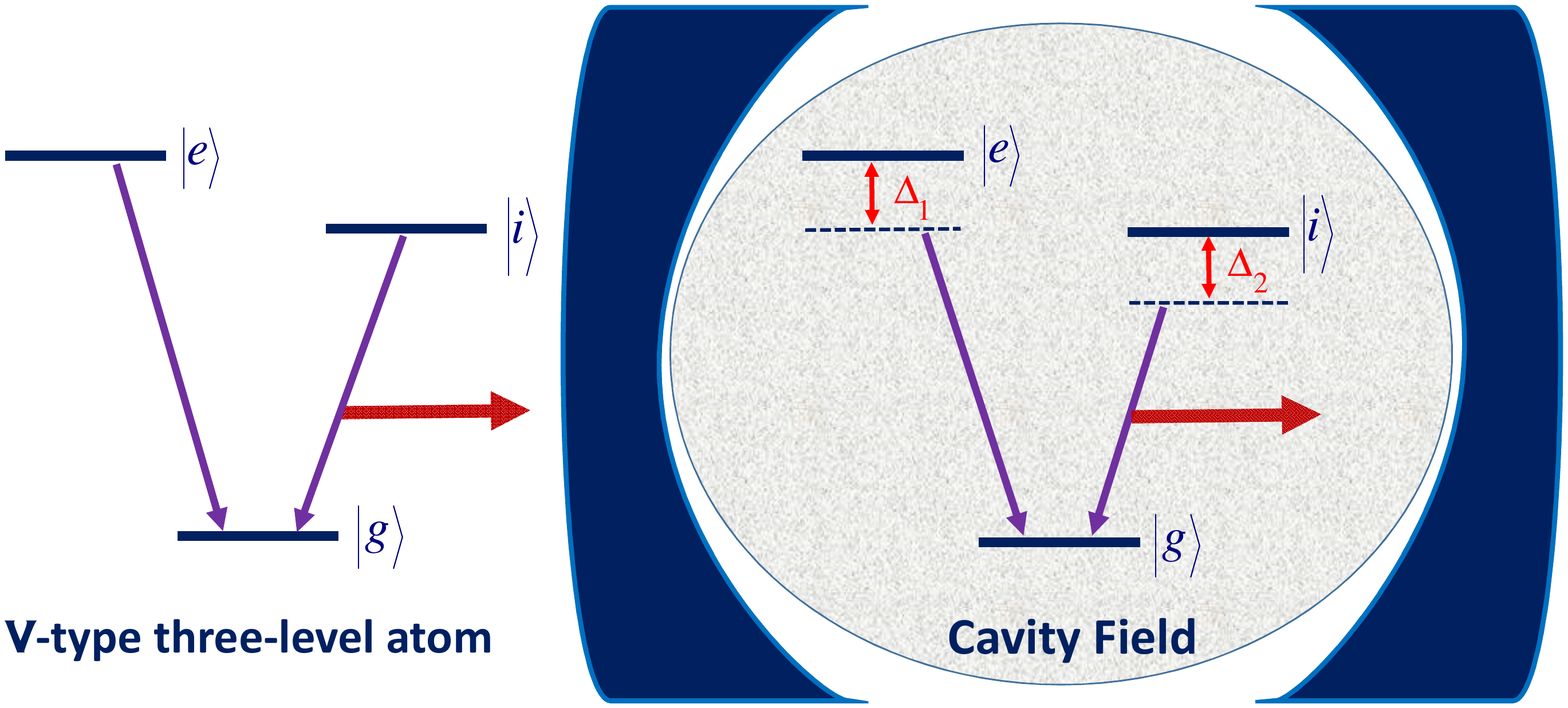}}
                                  \vspace {-1 cm}
 \caption{Schematic diagram of the interaction between two V-type three-level atoms passing consecutively through a cavity field.}
 \label{Diagram}
                              \end{figure}

   We consider two identical V-type three-level atoms passing through a cavity containing a  single-mode quantized field with frequency $ \nu $ \cite{Arpita}. We should emphasize that in this model, at any time there exists only one atom in the cavity. Therefore, passing the first atom through the cavity evolves the atom and the initial cavity field, from which one obtains particular circumstances for the atom and field that determine the initial field state for entering the second atom (after doing a measurement on the atom). Moreover, in general the atoms which respectively transit in the cavity interact  with the field non-resonantly and we suppose the intensity-dependent coupling regime. Also, since in general the V-type three-level atoms coupled to a field with $ \lambda_{1}\neq \lambda_{2} $ where $ \lambda_{1} $ and $ \lambda_{2} $ are the atom-field coupling constants. We do not consider equal coupling as in Ref. \cite{Arpita} has been done.
      Let us express the configuration of the atoms in detail. Their energy states are $ \omega_{e} $, $ \omega_{i} $ and $ \omega_{g} $  which correspond to the atomic energy levels $|e\rangle  $, $ |i\rangle $ and $ |g\rangle $ (Fig. \ref{Diagram}). The allowed  photon transitions  are $ |i\rangle  \longleftrightarrow  |g\rangle $ and $ |e\rangle \longleftrightarrow |g\rangle $, and the transition $ |i\rangle \longleftrightarrow  |e\rangle $ is forbidden.
       Since at any time there exists only one atom in the cavity, so the Hamiltonian of the atom-field system is the same at all times.
       Therefore, as the first step, we perform the Hamiltonian for the interacting subsystems by extending the JCM $ (\hbar=1) $  as follows \cite{Scully}:
   \begin{eqnarray}
   \hat{H} =\hat{H}_{0}+\hat{H}_{int},
   \end{eqnarray}
   where $ \hat{H}_{0} $  denotes the atom and field Hamiltonians in the absence of any interaction:
   \begin{eqnarray}
   \hat{H}_{0}=\nu\hat{a}^{\dag} \hat{a}+\omega_{e}|e\rangle \langle e|+\omega_{i} |i\rangle \langle i|+\omega_{g} |g\rangle \langle g|,
   \end{eqnarray}
   and $ \hat{H}_{int} $ is the interaction Hamiltonian in the rotating wave approximation:
    \begin{eqnarray}\label{Hint}
    \hat{H}_{int}&=&   \lambda_{1} (\hat{R}  |e\rangle \langle g|+  \hat{R}^{\dag} |g\rangle \langle e|)+\lambda_{2}(\hat{R} |i\rangle \langle g|+\hat{R}^{\dag}|g\rangle \langle i|),
    \end{eqnarray}
    which its form in the interaction picture reads as:
    \begin{eqnarray}\label{Hintint}
    \hat{V}_{I}&=&   \lambda_{1} (\hat{R}  e^{i\Delta_{1}t} |e\rangle \langle g|+  \hat{R}^{\dag}  e^{-i\Delta_{1}t} |g\rangle \langle e|)\nonumber\\
    &+&\lambda_{2}(\hat{R} e^{i\Delta_{2}t}|i\rangle \langle g|+\hat{R}^{\dag}e^{-i\Delta_{2}t}|g\rangle \langle i|),
    \end{eqnarray}
    where $ \Delta_{1}=(\omega_{e}-\omega_{g})-\nu $ and $ \Delta_{2}=(\omega_{i}-\omega_{g})-\nu $ are the detuning parameters. The operators $ \hat{a} $ and $ \hat{a}^{\dag} $ are the well-known bosonic annihilation and creation operators, respectively and two nonlinear operators $ \hat{R}=\hat{a}f(\hat{n}) $ and $ \hat{R}^{\dag} $=$ f(\hat{n})  \hat{a}^{\dag} $  satisfy the non-canonical commutation relation:
        \begin{eqnarray}
        [\hat{R}, \hat{R}^{\dag} ]= (\hat{n}+1)f^{2}(\hat{n}+1)-\hat{n}f^{2}(\hat{n}).
        \end{eqnarray}
     $ \hat{n}=\hat{a}^{\dagger}\hat{a} $ where $ \hat{n} $ is the number operator. $ \lambda_{1} $ , $ \lambda_{2} $ in (\ref{Hint}), (\ref{Hintint}) are the atom-field coupling constants; which in our nonlinear JCM they are changed to the intensity-dependent  atom-field coupling $ \lambda_{i}f(n) $ (with $ i=1,2) $  where $ f(n) $ represents an arbitrary nonlinearity function.
    To study the dynamics of the considered system, we should acquire its wave function at first. The general wave function for our system at any time $ t>0 $ can be written as a proper combination of the basic eigenstates of the atom and field, i.e.:
    \begin{eqnarray}\label{wave1}
    |\psi(t)\rangle&=&\sum_{n=0}^{+\infty}A_{1}(n,t)|e,n\rangle+B_{1}(n,t)|i,n\rangle+C_{1}(n+1,t)|g,n+1\rangle.
    \end{eqnarray}
        In order to obtain the probability amplitudes in $ |\psi(t)\rangle $, one should solve the time-dependent Schr\"{o}dinger equation $i \frac{\partial}{\partial t} |\psi(t)\rangle =\hat{V}_{I}| \psi(t)\rangle $. Along doing this task, one arrives at three coupled differential equations  in term of the above expansion coefficients in (\ref{wave1}):
        \begin{eqnarray}\label{diff eq1}
        i\;\dot{A_{1}}&=& \lambda_{1}\sqrt{n+1}f(n+1)e^{i\Delta_{1}t}C_{1}, \nonumber \\
        i\;\dot{B_{1}}&=& \lambda_{2}\sqrt{n+1}f(n+1)e^{i\Delta_{2}t}C_{1}, \nonumber \\
        i\;\dot{C_{1}}&=& \lambda_{1}\sqrt{n+1}f(n+1)e^{-i\Delta_{1}t}A_{1}\nonumber \\
        &+&\lambda_{2}\sqrt{n+1}f(n+1)e^{-i\Delta_{2}t}B_{1},
        \end{eqnarray}
        Now, by considering $ B_{1}=e^{i\mu t} $ and inserting it into these equations, an algebraic third-order equation in the following form is obtained:
        \begin{eqnarray}\label{three eq11}
        \mu^{3}+x_{1}\mu^{2}+x_{2}\mu+x_{3}=0,
        \end{eqnarray}
        where
        \begin{eqnarray}
        x_{1}&=&\Delta_{1}-2 \Delta_{2}, \nonumber \\
        x_{2}&=&\Delta_{2}^{2}-\Delta_{2}\Delta_{1}-(\lambda_{1}^{2}+\lambda_{2}^{2})(n+1)f^{2}(n+1), \nonumber \\
        x_{3}&=&\lambda_{2}^{2}(\Delta_{2}-\Delta_{1})(n+1)f^{2}(n+1).
        \end{eqnarray}
      Three different roots of (\ref{three eq11}) are as follow \cite{shu1,kardan}:
          \begin{eqnarray}\label{roots}
        \mu_{r}&=&-\frac{1}{3}x_{1}+\frac{2}{3}\sqrt{x^{2}_{1}-3x_{2}}\cos\left[ \theta+\frac{2}{3}(r-1)\pi \right],\;\;\;\;\;r=1,2,3. \nonumber
        \end{eqnarray}
        with
        \begin{eqnarray}
        \theta &=& \frac{1}{3}\cos^{-1}\left[ \frac{9x_{1}x_{2}-2x^{3}_{1}-27 x_{3}}{2(x^{2}_{1}-3x_{2})^{3/2}}\right].
        \end{eqnarray}
        Consequently, $  B_{1} $ can be written as a linear combination of $ e^{i\mu_{j}t} $ in the form:
        \begin{eqnarray}\label{b}
        B_{1}=\sum_{j=1}^{3}k_{j}e^{i\mu_{j}t}.
        \end{eqnarray}
        Finally, by replacing equation (\ref{b}) in (\ref{diff eq1}) and after some lengthy but straightforward manipulations, we arrive at explicit form of the probability amplitudes in (\ref{wave1}) as follow:
    \begin{eqnarray}\label{zarayeb1}
    A_{1}(n,t)&=&\sum_{j=1}^{3}\frac{k_{j}e^{i(\mu_{j}-\Delta_{2}+\Delta_{1})t}(\mu_{j}^{2}-\Delta_{2}\mu_{j} -\lambda_{2}^{2}(n+1)f^{2}(n+1)) }{\lambda_{1}\lambda_{2}(n+1)f^{2}(n+1)},\nonumber\\
    B_{1}(n,t)&=&\sum_{j=1}^{3} k_{j}e^{i\mu_{j} t}, \nonumber\\
    C_{1}(n+1,t)&=&\sum_{j=1}^{3}\frac{-k_{j}\mu_{j}e^{i(\mu_{j}-\Delta_{2})t}}{\lambda_{2}\sqrt{n+1}f(n+1)}.
    \end{eqnarray}
    The coefficients $k_j$ can be determined by the initial conditions of atom and field. For this purpose, we suppose that initially the first atom is in a coherent superposition of two of its excited states, i.e.,:
    \begin{eqnarray}\label{sha1}
    |\psi(0)\rangle_{A}&=&\frac{1}{\sqrt{2}}(|e\rangle+|i\rangle),
    \end{eqnarray}
    and the field is prepared in the coherent state:
    \begin{eqnarray}\label{shf}
    |\psi(0)\rangle_{F}= \sum_{n=0}^{+\infty} F_{n}  |n\rangle,\;\;\;\;\;\;\;\; F_{n} = e^{-\frac{\bar{n}}{2}}\frac{\alpha^{n}}{\sqrt{n!}},
    \end{eqnarray}
    where $\bar{n}=|\alpha|^{2} $ implies the initial average photon number of the field. Now, by replacing (\ref{sha1}) and (\ref{shf}) in the wave function (\ref{wave1}) with the introduced amplitudes in (\ref{zarayeb1}), we can derive:
   \begin{eqnarray}
      k_{j}&=&\frac{F_{n}}{\sqrt{2}\mu_{jl}\mu_{jk}}\big(\mu_{k}\mu_{l}+(\lambda_{2}^{2}+\lambda_{1}\lambda_{2})(n+1)f^{2}(n+1)\big),
      \end{eqnarray}
    where $ \mu_{jk}=\mu_{j}-\mu_{k} $ and $ j\neq k\neq l=1,2,3 $. $ A_{1}(n,t),B_{1}(n,t)  $ and $ C_{1}(n+1,t) $ are complex values that satisfy the normalization condition $\sum_{n=0}^{+\infty}|A_{1}(n,t)|^{2}+|B_{1}(n,t)|^{2}+|C_{1}(n+1,t)|^{2}=1 $. Consequently, the state vector of the considered atom-field system at any time $ t>0 $ is obtained completely.
    Now, we assume that the atom is detected in its ground state, after the interaction time $t=t_{1}$ between atom and the cavity field. In our case, this particular time may be appropriately chosen using the dynamics of the population inversion. By this time, we will determine the initial field condition for entering the second atom and allowing it to interact with the field. Accordingly, the normalized state of the field, after the projection of the atom into the state $|g\rangle$, reads as\footnote{In Ref. \cite{Arpita} the author did not explicitly refer to the normalization condition of the final field state after projecting the atom-field state into its ground state $ |g\rangle $ in the first atom-field interaction procedure.}:
    \begin{eqnarray}\label{fieldstate2}
    |\psi_{1}(t_{1})\rangle_F&=& \Big(\sum_{n=0}^{+\infty}|C_{1}(n+1,t_{1})|^{2}\Big)^{-\frac{1}{2}}\sum_{n=0}^{+\infty}C_{1}(n+1,t_{1}) |n+1\rangle,
    \end{eqnarray}
    where $ C_{1}(n+1,t_{1}) $  is determined by the third relation of (\ref{zarayeb1}). At this moment, we allow the second atom enters the cavity while it is in a coherent superposition state $  |\psi_{2}(t_{2}=0)\rangle_{A}=\frac{1}{\sqrt{2}}(|e\rangle+|i\rangle) $, like the first atom.  The atom-field interaction occurs in the time interval $ t_2 $; briefly $  |\psi_{2}(t_{1};t_{2}=0)\rangle_{F} =  |\psi_{1}(t_{1})\rangle_{F} $. Generally, similar to (\ref{wave1}) we suppose that the state of the system along the passage of the second atom changes to:
    \begin{eqnarray}\label{wave2}
    |\psi_{2}(t_{1},t_{2})\rangle&=&\sum_{n=0}^{+\infty}A_{2}(n,t_{1},t_{2})|e,n\rangle+B_{2}(n,t_{1},t_{2})|i,n\rangle\nonumber\\
   &+&C_{2}(n+1,t_{1},t_{2})|g,n+1\rangle.
    \end{eqnarray}
    Again, with the help of the time-dependent Schr\"{o}dinger equation, the following coupled differential equations are obtained:
    \begin{eqnarray}\label{diff eq}
    i\;\dot{A_{2}}&=& \lambda_{1}\sqrt{n+1}f(n+1)e^{i\Delta_{1}t_{2}}C_{2}, \nonumber \\
    i\;\dot{B_{2}}&=& \lambda_{2}\sqrt{n+1}f(n+1)e^{i\Delta_{2}t_{2}}C_{2}, \nonumber \\
    i\;\dot{C_{2}}&=& \lambda_{1}\sqrt{n+1}f(n+1)e^{-i\Delta_{1}t_{2}}A_{2}\nonumber \\
    &+&\lambda_{2}\sqrt{n+1}f(n+1)e^{-i\Delta_{2}t_{2}}B_{2},
    \end{eqnarray}
    where the dot refers to differentiation with respect to time $ t_{2} $. Now, if we set $ B_{2}=e^{iut_{2}} $ and by using the relations in (\ref{diff eq})  we arrive at an algebraic third-order equation:
    \begin{eqnarray}\label{three eq1}
    u^{3}+y_{1}u^{2}+y_{2}u+y_{3}=0,
    \end{eqnarray}
    where
    \begin{eqnarray}
    y_{1}&=&\Delta_{1}-2 \Delta_{2}, \nonumber \\
    y_{2}&=&\Delta_{2}^{2}-\Delta_{2}\Delta_{1}-(\lambda_{1}^{2}+\lambda_{2}^{2})(n+1)f^{2}(n+1), \nonumber \\
    y_{3}&=&\lambda_{2}^{2}(\Delta_{2}-\Delta_{1})(n+1)f^{2}(n+1).
    \end{eqnarray}
  Three different roots of (\ref{three eq1}) can be determined like the equations (\ref{three eq11}) and(\ref{roots}). So, the general form of $  B_{2} $  reads as the linear combination of $ e^{iu_{r}t_{2}} $ as $B_{2}=\sum_{r=1}^{3}q_{r}e^{iu_{r}t_{2}} $. Replacing this summation into the relations (\ref{diff eq}) and after some lengthy manipulations, finally arrive us at the explicit form of the expansion coefficients of our final atom-field state vector (\ref{wave2}) as below:
    \begin{eqnarray}\label{coefficients2}
    A_{2}(n,t_{1},t_{2})&=&\sum_{r=1}^{3}\frac{q_{r}e^{i(u_{r}-\Delta_{2}+\Delta_{1})t_{2}}}{\lambda_{1}\lambda_{2}(n+1)f^{2}(n+1)}\nonumber\\
   &\times& (u^{2}_{r}-\Delta_{2}u_{r}-\lambda_{2}^{2} (n+1)f^{2}(n+1)),\nonumber\\
    B_{2}(n,t_{1},t_{2})&=&\sum_{r=1}^{3} q_{r}e^{iu_{r} t_{2}}, \nonumber\\
    C_{2}(n+1,t_{1},t_{2})&=&\sum_{r=1}^{3}\frac{-q_{r}u_{r}e^{i(u_{r}-\Delta_{2})t_{2}}}{\lambda_{2}\sqrt{n+1}f(n+1)},
    \end{eqnarray}
    where the coefficients $q_{r}$ should be determined via the initial condition of the second atom and field in (\ref{fieldstate2}). Then, by using (\ref{coefficients2}) the following relations may be obtained:
    \begin{eqnarray}
    q_{r}&=&\frac{C_{1}(n+1,t_{1})}{\sqrt{2\sum_{n=0}^{+\infty}|C_{1}(n+1,t_{1})|^{2}}}\nonumber\\
    &\times&\Big(\frac{(\lambda_{2}^{2}+\lambda_{1}\lambda_{2})(n+1)f^{2}(n+1)+u_{l}u_{k}}{u_{rl}u_{rk}}\Big),
    \end{eqnarray}
    where $ u_{rl}=u_{r}-u_{l} $ and $ r\neq k\neq l=1,2,3 $. Accordingly, the wave function of the entire atom-field system $ |\psi_{2}(t_{1},t_{2})\rangle $ in (\ref{wave2}) is completely determined in an explicit form. Finding the whole considered system state, allows us to evaluate all its physical properties.
   \section{Physical properties}\label{phyprop}
 Now, which we have obtained the explicit form of the final atom-field state vector, we are able to analyze its physical properties. However,
 in order to simplify our presentation, we can transform the results in the previous section such that, all necessary quantities can be
 computed if we determine the relative values of  $ \lambda_{1}/\lambda_{2} $ and  $ \Delta_{i}/\lambda_{2}; i=1,2 $.
 We consider $ \lambda_{1}/\lambda_{2} =0.9 $ in all of the numerical calculations in the remainder of this paper.
  Consequently, we can plot all required quantities as function of the scaled times $ \tau_{1}=\lambda_{2}t_{1} $ and
  $ \tau_{2}=\lambda_{2}t_{2}  $ for the interaction times of  the first and second atom, respectively. It is also worth mentioning that we  present our numerical results, by choosing the well-known nonlinearity function $f(n)=\sqrt{n} $
  as the intensity-dependent function. In addition, (i) due to the minor correction applied on \cite {Arpita},
  in addition to the fact that (ii) we will consider the off-resonant case and (iii) nonequal constant couplings
  $(\lambda_1\neq \lambda_2)$, we also plot the related figures for $f(n)=1$, too.
                 \subsection{The population inversion}\label{Population}

\begin{figure}[htbp]
                            \centerline{\includegraphics[width=0.95\columnwidth]{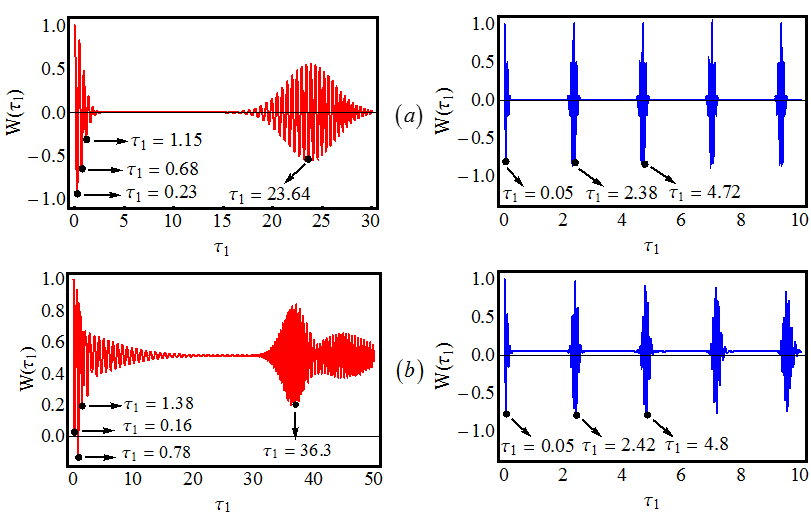}}
 \caption{The atomic inversion for the first atom as a function of the scaled time $\tau_{1}=\lambda_{2} t_{1}$  when the atom is initially in a coherent superposition of their excited states and the field is in a coherent state with $|\alpha|^{2} = 25 $. The left plots correspond to the absence of the intensity-dependent coupling $f(n)=1$, and the right ones are plotted in the presence of the intensity-dependent coupling with nonlinearity function $f(n) = \sqrt{n}$. Also,
   (a)  $\Delta_{1}=\Delta_{2}=0$, (b) $\frac{\Delta_{1}}{\lambda_{2}}=7,\frac{\Delta_{2}}{\lambda_{2}}=15$.}
 \label{populationinversionI}
                         \end{figure}

\begin{figure}[htbp]
   \centerline{\includegraphics[width=0.95\columnwidth]{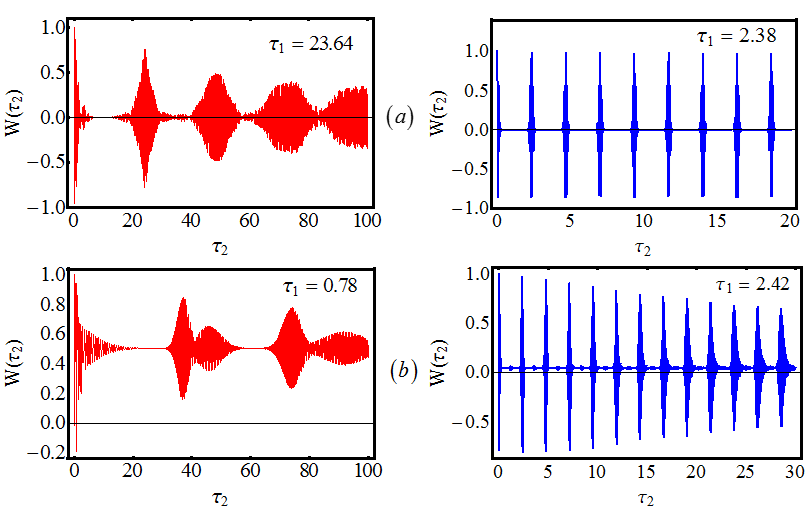}}
 \caption{The atomic inversion when the second atom entered the cavity as a function of  the scaled time $\tau_{2}=\lambda_{2} t_{2}$ for particular chosen values of the scaled time $ \tau_{1}=\lambda_{2}t_{1} $ extracted from Fig. \ref{populationinversionI},  when the atom is initially in a coherent superposition of its excited states. The left plots correspond to the absence of the intensity-dependent coupling $f(n)=1$, and the right ones are plotted in the presence of the intensity-dependent coupling with nonlinearity function $f(n) = \sqrt{n}$. Also,
    (a)  $\Delta_{1}=\Delta_{2}=0$, (b) $\frac{\Delta_{1}}{\lambda_{2}}=7,\frac{\Delta_{2}}{\lambda_{2}}=15$.}
 \label{populationinversionII}
  \end{figure}

 The atomic population inversion is defined as  a measure of energy exchange between atom and field. Studying  this quantity in the full quantum mechanical approach is usually together with appearing the collapse and revival phenomena that are resulted from the discrete nature of photons (field quantization). The so-called population inversion for a V-type three-level atom can be  computed by the difference between the probabilities of finding the atom in the two excited states and the ground state, i.e., $ W(t_{2})=(\rho_{ee}+\rho_{ii})-\rho_{gg}$.\\
 When the first atom passes through the cavity, the values of the probability amplitudes of the atom-field system in the cavity is in term of the scaled time $ \tau_{1} $ may be derived,  from which one can obtain the temporal behavior of the atomic inversion for the first atom ($ W(\tau_{1}) $). We considered the field state (\ref{fieldstate2}) as initial field condition for passing the second atom, i.e., the atomic state $ |g\rangle $  is the result of the atomic measurement. We are going to plot the atomic inversion after  passing only the first atom to achieving such a condition appropriately (see Fig. \ref{populationinversionI}). To state more explicitly, when (the scaled times which) the atomic inversion is in its minima values the atomic state may be observed in the ground state of the V-type atom (we will need these scaled times for our further calculations in this paper).
From Fig. \ref{populationinversionI} we can choose the appropriate moments of the scaled time $ \tau_{1} $ for which the above-mentioned condition may properly happen. Thus, we are able to use the final state vector (after entering the second atom) (Eq. (\ref{coefficients2})), correctly.  In Fig. \ref{populationinversionII} we plotted the atomic inversion for the second atom in term of the scaled time $ \tau_{2} $ for particular chosen values of the scaled time $ \tau_{1} $ extracted from Fig. \ref{populationinversionI}.  The left figures show the numerical results in the absence of the intensity-dependent coupling  and the right ones deal with the presence of the intensity-dependent coupling. The above two plots
$ \ref{populationinversionII}(a) $ indicate the exact resonance case $ (\Delta_{1}=\Delta_{2}=0) $ and the two below plots $ \ref{populationinversionII}(b) $ display the effect of the detuning parameters $ (\frac{\Delta_{1}}{\lambda_{2}}=7 ,\frac{\Delta_{2}}{\lambda_{2}}=15 )$. The initial mean number of photons of the field is considered as $ |\alpha|^{2}=25 $. In all plotted curves in Fig. $ \ref{populationinversionII}$ one can see that the atomic population inversion is clearly occurred together with the collapse-revival phenomena.
    From the plotted curves one  may conclude that the nonlinearity in the atom-field coupling makes the patterns of collapse-revival more visible, in particular nearly the full revival is occurred in the resonance condition and intensity-dependent coupling.
                        Moreover, in the absence of nonlinearity, the detuning causes a noticeable shift to positive values of atomic inversion in the pattern of collapse-revival, while  the revival amplitudes in the nonlinear regime experience slightly decrease by entering the detuning parameters.
   \subsection{The von Neumann entropy}\label{Von}
\begin{figure}[htbp]
           \centerline{\includegraphics[width=0.95\columnwidth]{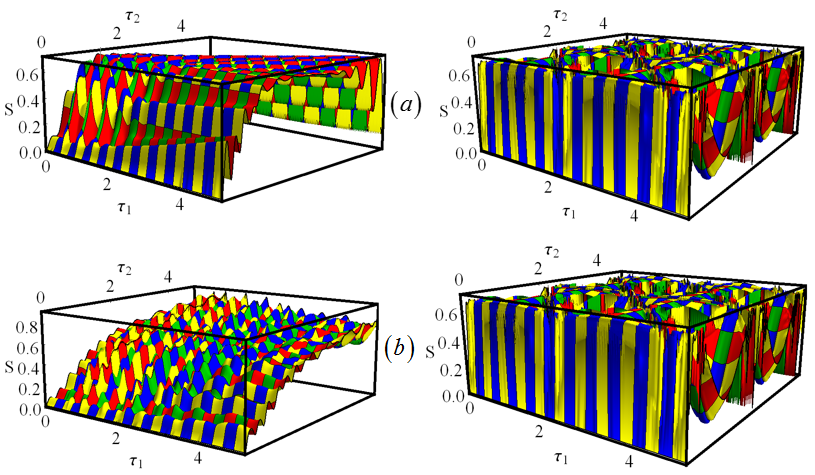}}
           \caption{ Three-dimensional plots of the time evolution of the von Neumann entropy in term of the scaled times $ \tau_{1}=\lambda_{2}t_{1} $ and $ \tau_{2}=\lambda_{2}t_{2} $; other parameters are chosen similar to Fig. \ref{populationinversionII}.}
          \label{vonNeumannEntropy}
           \end{figure}

      Entanglement is a pure quantum phenomenon that Schr\"{o}dinger designed it as a private trait of quantum mechanic. It shows the nonclassical correlation in the information aspects. Entanglement may be quantified via the evaluation of the entropy which represents a measure of lack of  information from the system \cite{knightbk}. Actually, the field entropy is a criterion that shows the degree of entanglement. In this regard, Araki-Lieb theorem \cite{araki1} demonstrates that, by starting from the initial pure state of any atom-field system, one has $ S_{A}(t)=S_{F}(t) $ at any time $ t>0 $ \cite{araki2}(notice that in our model we have $ S_{A(F)}(t_{1},t_{2}) $ instead of $ S_{A(F)}(t) $, where  $ S_{A(F)}(t_{1},t_{2}) $ denote the entropy of the second atom(field)).  Notice that the effect of passing the first atom in the state of the system has been taken into account in selecting particular form of the initial field state. So because of the mentioned equivalence, we can calculate either the atomic or field entropy. We use the von Neumann entropy which  is defined in terms of the  reduced density matrix (of the second atom)  as \cite{pk3}:
    \begin{eqnarray}
    S_{A(F)}(t_{1},t_{2})=-\mathrm{Tr_{A(F)}(\rho_{A(F)}\ln\rho_{A(F)}  )}
    \end{eqnarray}
    where
       \begin{eqnarray}\label{matrix}
       \hat{\rho}_{A}(t_{1},t_{2})&=&\mathrm{Tr}_{F}\left(  |\psi_{2}(t_{1},t_{2}) \rangle \langle \psi_{2}(t_{1},t_{2}) |   \right)  \nonumber \\
       &=&\left(
       \begin{array}{ccc}
       \rho_{ee} & \rho_{ei} & \rho_{eg} \\
       \rho_{ie} & \rho_{ii} & \rho_{ig} \\
       \rho_{ge} & \rho_{gi} & \rho_{gg} \\
       \end{array}
       \right).
       \end{eqnarray}
       The above matrix elements  are explicitly given as below:
       \begin{eqnarray}\label{atomic1}
       \rho_{ee}&=&\sum_{n=0}^{+\infty}  A_{2}(n,t_{1},t_{2}) A_{2}^{*}(n,t_{1},t_{2}), \nonumber \\
       \rho_{ei}&=& \rho_{ie}^{*}= \sum_{n=0}^{+\infty}  A_{2}(n,t_{1},t_{2}) B_{2}^{*}(n,t_{1},t_{2}), \nonumber \\
       \rho_{eg}&=&\rho_{ge}^{*}=\sum_{n=0}^{+\infty}  A_{2}(n+1,t_{1},t_{2}) C_{2}^{*}(n+1,t_{1},t_{2}), \nonumber \\
       \rho_{ii}&=&\sum_{n=0}^{+\infty}  B_{2}(n,t_{1},t_{2}) B_{2}^{*}(n,t_{1},t_{2}), \nonumber \\
       \rho_{ig}&=&\rho_{gi}^{*}=\sum_{n=0}^{+\infty}  B_{2}(n+1,t_{1},t_{2}) C_{2}^{*}(n+1,t_{1},t_{2}), \nonumber \\
       \rho_{gg}&=&\sum_{n=0}^{+\infty}  C_{2}(n+1,t_{1},t_{2}) C_{2}^{*}(n+1,t_{1},t_{2}).
       \end{eqnarray}
       $  A_{2}(n,t_{1},t_{2}),  B_{2}(n,t_{1},t_{2}) $ and $ C_{2}(n+1,t_{1},t_{2}) $      are the amplitudes
       which have been determined in section (\ref{sec-intro}). Meanwhile,  the von Neumann entropy of the system can be obtained as follows \cite{pk3}:
      \begin{eqnarray}
    S_{F}(t_{1},t_{2})=S_{A}(t_{1},t_{2})=-\sum_{j=1}^{3}\gamma_{j} \ln \gamma_{j},
    \end{eqnarray}
    where $ \gamma_{j} $ represents the eigenvalues of the reduced density matrix of the atom in (\ref{matrix}) may be expressed as:
    \begin{eqnarray}
    \gamma_{j}&=&-\frac{1}{3}\beta_{1}+\frac{2}{3}\sqrt{\beta^{2}_{1}-3\beta_{2}}\cos\left[\alpha+\frac{2}{3}(j-1)\pi \right], \nonumber \\
    \alpha &=& \frac{1}{3}\cos^{-1}\left[ \frac{9\beta_{1}\beta_{2}-2\beta^{3}_{1}-27\beta_{3}}{2(\beta^{2}_{1}-3\beta_{2})^{3/2}}\right],
    \end{eqnarray}
    with
    \begin{eqnarray}\label{vzal}
    \beta_{1}&=&-\rho_{ee}-\rho_{ii}-\rho_{gg}=-1, \nonumber \\
    \beta_{2}&=&\rho_{ee}\rho_{ii}+\rho_{ii}\rho_{gg}+\rho_{gg}\rho_{ee} -\rho_{ei}\rho_{ie}-\rho_{ig}\rho_{gi}-\rho_{ge}\rho_{eg}, \nonumber \\
    \beta_{3}&=& -\rho_{ee}\rho_{ii}\rho_{gg}-\rho_{ei}\rho_{ig}\rho_{ge}-\rho_{eg}\rho_{gi}\rho_{ie}+\rho_{ee}\rho_{ig}\rho_{gi}\nonumber\\ &+&\rho_{ii}\rho_{ge}\rho_{eg}+\rho_{gg}\rho_{ei}\rho_{ie}.
    \end{eqnarray}
     Now, we are ready to present the evolution of entropy. The three-dimensional Fig. \ref{vonNeumannEntropy} represents the dynamics of the von Neumann entropy with respect to the scaled times $ \tau_{1}$ and $ \tau_{2} $.
 It is readily observed that,  the four plotted figures show the notable entanglement between atom and field. By comparing the left plots of  $ \ref{vonNeumannEntropy}(a) $ and $ \ref{vonNeumannEntropy}(b) $, i.e., the graphs for the atom-field constant coupling, one can realize that entering the detuning parameters increases the amount of entropy (and so the entanglement measure), while paying attention to the right graphs, i.e, in the nonlinear regime, arrives one to the fact that the detuning does not have significant effect on the amount of entropy.
%

%
         \subsection{The field squeezing}\label{Squeezing1}
\begin{figure}[htbp]
   \centerline{\includegraphics[width=0.95\columnwidth]{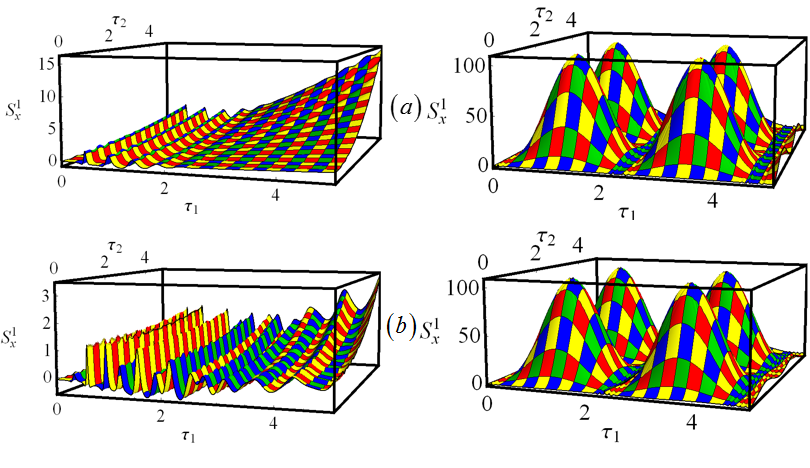}}
 \caption{Three-dimensional plots of the time evolution of normal squeezing of the field in term of the scaled times $ \tau_{1}=\lambda_{2}t_{1} $ and $ \tau_{2}=\lambda_{2}t_{2} $; other parameters are chosen similar to Fig. \ref{populationinversionII}.}
   \label{sx1}
  \end{figure}

\begin{figure}[htbp]
   \centerline{\includegraphics[width=0.95\columnwidth]{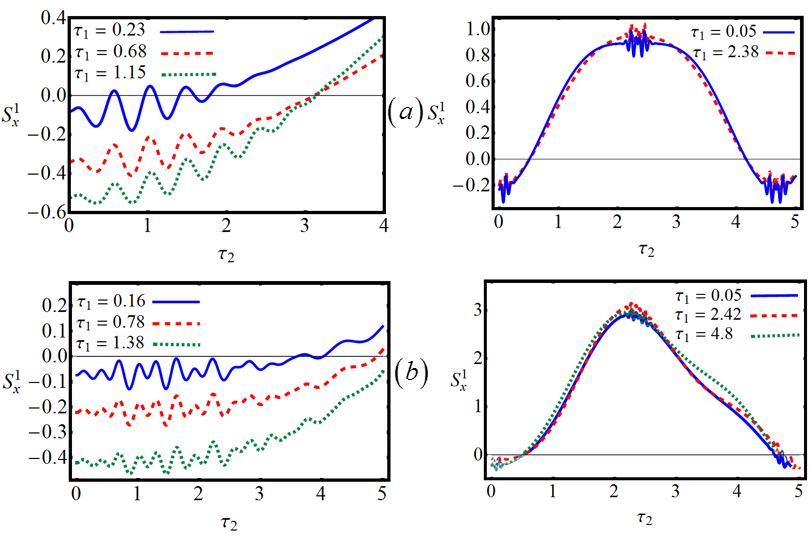}}
   \caption{The cross-section of the plots of Fig. \ref{sx1}: normal squeezing as a function of the scaled time $   \tau_{2}=\lambda_{2}t_{2} $ for particular values of the scaled time $\tau_{1}=\lambda_{2}t_{1} $ in Fig. \ref{populationinversionI}; other parameters are chosen similar to Fig. \ref{populationinversionII}.}
 \label{sx12}
 \end{figure}

\begin{figure}[htbp]
   \centerline{\includegraphics[width=0.95\columnwidth]{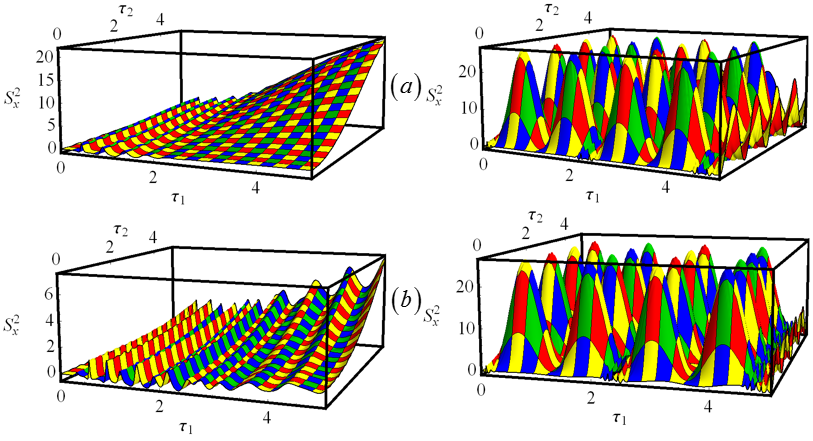}}
   \caption{Three-dimensional plots of the time evolution of the second-order squeezing of the field in term of the scaled times $ \tau_{1}=\lambda_{2}t_{1} $ and $ \tau_{2}=\lambda_{2}t_{2} $; other parameters are chosen similar to Fig. \ref{populationinversionII}.}
          \label{sx2}
   \end{figure}

\begin{figure}[htbp]
   \centerline{\includegraphics[width=0.95\columnwidth]{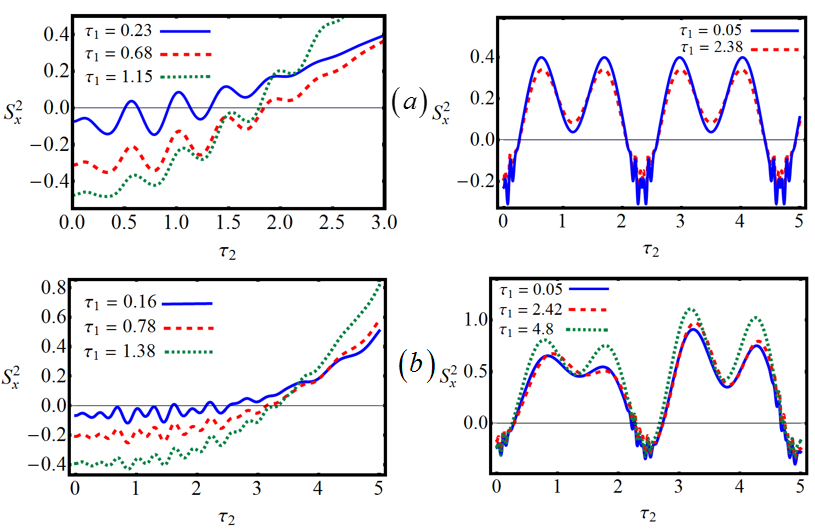}}
  \caption{The cross-section of the plots of Fig. \ref{sx2}: the second-order squeezing as a function of the scaled time $ \tau_{2}=\lambda_{2}t_{2}$ for particular values of the scaled time $\tau_{1}=\lambda_{2}t_{1} $ in Fig. \ref{populationinversionI}; other parameters are chosen similar to Fig. \ref{populationinversionII}. }
   \label{sx22}
 \end{figure}

       It is well known that the nonclassical light is a radiation field that it does not have any classical analogue.
        Squeezing is another  important nonclassical phenomenon in the framework of quantum optics.
        It is described with reducing the noise in one of  quadratures of the field  in comparison to the coherent state or the vacuum state with the price of an increase of the noise in the other quadrature of the field such that  the uncertainly relation is still satisfied. Supposing  the  two operators $\hat{x}$ and $\hat{y}$  possess the  commutation relation $ [\hat{x},\hat{y}]=\hat{z} $, then the uncertainly relation is written as $ \Delta \hat{x}\Delta \hat{y}\geqslant\frac{1}{2}|\langle \hat{z}\rangle| $ where $ \Delta \hat{s}= \sqrt{\langle\hat{s}^{2}\rangle-\langle \hat{s}\rangle^{2}}$. For evaluating this quantity, we define two quadrature components of the field in terms of the bosonic operators $ \hat{x}_{k}=(\hat{a}^{k}+(\hat {a}^{\dag})^{k})/\sqrt{2} $ and $ \hat{p}_{k}=({{\hat a}^k}-{(\hat a^\dag)^k})/\sqrt{2}i $ where the subscript $ k=1,2,3,... $ shows the order of squeezing of the electromagnetic field.  In these relations the first-order and the second-order squeezing correspond respectively to $ k=1 $ and $ k=2 $. In order to calculate the normal squeezing of the field, one can obtain simply $ [x_{1},p_{1}]=i $ with  the uncertainly relation  $ (\Delta \hat{x})^{2}(\Delta \hat{p})^{2}\geq\frac{1}{4} $.  Equivalently, we can consider the quadratures variances by defining the following squeezing parameters $ S_{x}^{(1)}=((\Delta \hat{x})^{2}-0.5)/0.5 $ and $ S_{p}^{(1)}=((\Delta \hat{p})^{2}-0.5)/0.5 $. Consequently, squeezing happens if $ -1<S_{x}^{(1)}<0 $ or $ -1<S_{p}^{(1)}<0 $. The above relations can be rewritten as:
        \begin{eqnarray}
        S_{x}^{(1)}&=&\langle\hat{a}^{2}\rangle+\langle\hat{a}^{\dag^{2}}\rangle+2\langle\hat{a}^{\dag}\hat{a}\rangle-2\langle \hat{a}\rangle\langle\hat{a}^{\dag}\rangle-\langle\hat{a}\rangle^{2}-\langle\hat{a}^{\dag}\rangle^{2},\ \nonumber \\
        S_{p}^{(1)}&=&-\langle\hat{a}^{2}\rangle-\langle\hat{a}^{\dag^{2}}\rangle+2\langle\hat{a}^{\dag}\hat{a}\rangle-2\langle \hat{a}\rangle\langle\hat{a}^{\dag}\rangle +\langle\hat{a}\rangle^{2}+\langle\hat{a}^{\dag}\rangle^{2}.\nonumber\\
        \end{eqnarray}
       To go further, we have to obtain the following expectation value with respect to the state of the system in (\ref{wave2}):
      \begin{eqnarray}\label{number}
      \langle\hat{a}^{\dag}\hat{a}\rangle&=&\sum_{n=0}^{+\infty}\Big(n(|A_{2}(n,t_{1},t_{2})|^{2}+|B_{2}(n,t_{1},t_{2})|^{2})\nonumber\\
      &+&(n+1)|C_{2}(n+1,t_{1},t_{2})|^{2}\Big),
      \end{eqnarray}
     \begin{eqnarray}\label{power}
     \langle\hat{a}^{r}\rangle&=&\sum_{n=0}^{+\infty}\sqrt{\frac{(n+r)!}{n!}}\Big[A_{2}(n+r,t_{1},t_{2}) A_{2}^{*}(n,t_{1},t_{2})\nonumber\\
     &+& B_{2}(n+r,t_{1},t_{2}) B_{2}^{*}(n,t_{1},t_{2})\nonumber\\
     &+& \sqrt{\frac{(n+r+1)}{(n+1)}}C_{2}(n+r+1,t_{1},t_{2}) C_{2}^{*}(n+1,t_{1},t_{2})\Big],
     \end{eqnarray}
     in which probability amplitudes $A_2, B_2$ and  $C_2$ were determined in (\ref{coefficients2})
          and clearly $ \langle\hat{a}^{\dagger^{r}}\rangle =\langle\hat{a}^{r}\rangle^{*} $.
          The three-dimensional plots in Fig.  \ref{sx1} show the behavior of $ S_{x}^{(1)} $ versus the scaled times $ \tau_{1} $ and $ \tau_{2} $  in which we have used the same parameters as in Fig. $ \ref{populationinversionII} $.
     Moreover, to increase our precision for the investigation of the occurrence of squeezing, we have depicted four two-dimensional plots in Fig.  \ref{sx12} in terms of the scaled time $\tau_{2}$, for particular the scaled times $\tau_{1}$ which are distinguished (and extracted) from Fig. $ \ref{populationinversionI} $ (in other words, these two-dimensional plots are indeed appropriate cross-sections of Fig. \ref{sx1}). In  both of resonance and nonresonance conditions and in the absence and presence of intensity-dependent atom-field coupling  (Figs. \ref{sx1} and \ref{sx12}, respectively), we see that squeezing takes place appropriately in  intervals of time at the beginnings of the interaction.\\
     Now, we turn our attention to the evaluation of the second-order squeezing ($k=2$). The operators of considered squeezing obey of the commutation relation $ [\hat{x}_{2},\hat{p}_{2}]=(2\hat{n}+1)i $ with  the uncertainly relation  $ (\Delta \hat{x}_{2})^{2}(\Delta \hat{p}_{2})^{2}\geqslant|\langle \hat{n}+\frac{1}{2}\rangle|^{2} $. Then the amplitude-squared squeezing parameters can be given as $ S_{x}^{(2)}=((\Delta \hat{x}_{2})^{2}-\langle \hat{n}+\frac{1}{2} \rangle)/\langle \hat{n}+\frac{1}{2} \rangle $ and $ S_{p}^{(2)}=((\Delta \hat{p}_{2})^{2}-\langle \hat{n}+\frac{1}{2} \rangle)/\langle \hat{n}+\frac{1}{2} \rangle $.
    One can rewrite the above relations as:
    \begin{eqnarray}
    S_{x}^{(2)}&=&\frac{1}{4\langle \hat{n} \rangle +2 }(\langle\hat{a}^{4}\rangle +\langle\hat{a}^{\dag^{4}}\rangle+2\langle ( \hat{a}^{\dag}\hat{a})^{2}\rangle-2\langle \hat{a}^{\dag} \hat{a}\rangle \nonumber \\
    &-&(\langle\hat{a}^{2}\rangle+\langle\hat{a}^{\dag^{2}}\rangle)^{2}),\nonumber \\
    S_{p}^{(2)}&=&\frac{1}{4\langle \hat{n} \rangle +2 }(2\langle ( \hat{a}^{\dag}\hat{a})^{2}\rangle-2\langle \hat{a}^{\dag} \hat{a}\rangle-\langle\hat{a}^{4}\rangle -\langle\hat{a}^{\dag^{4}}\rangle \nonumber\\
    &+&(\langle\hat{a}^{\dag^{2}}\rangle-\langle\hat{a}^{2}\rangle)^{2}).
    \end{eqnarray}
        To evaluate the second-order squeezing, we require the following mean value:
       \begin{eqnarray}\label{number2}
                 \langle(\hat{a}^{\dag}\hat{a})^{2}\rangle&=&\sum_{n=0}^{+\infty}\Big(n^{2}(|A_{2}(n,t_{1},t_{2})|^{2}+|B_{2}(n,t_{1},t_{2})|^{2})\nonumber\\
                 &+&(n+1)^{2}|C_{2}(n+1,t_{1},t_{2})|^{2}\Big),
                 \end{eqnarray}
                 where $A_2, B_2 $ and $ C_2$ were found in (\ref{coefficients2}).
       Now, using the relations (\ref{number}), (\ref{power}) and (\ref{number2}), we are able to calculate $ S_{x}^{(2)} $ and $ S_{p}^{(2)} $.
       The three-dimensional plots in Fig. \ref{sx2} indicate the second-order squeezing in the $ \hat{x}_{2} $ versus the scaled times $ \tau_{1} $ and $ \tau_{2} $.
       Also, plotted figures in Fig. \ref{sx22} are the cross-sections of the plots of Fig. \ref{sx2} in two-dimension in term of the scaled time $ \tau_{2} $ for a few particular
       the scaled times $ \tau_{1} $ (which are determined in Fig. $ \ref{populationinversionI} $).
         In the linear regime  (the left plots of Figs. $ \ref{sx2} $ and $ \ref{sx22} $)
                 squeezing is visible at the beginnings of the interaction in both of resonance and nonresonance conditions.
        In the intensity-dependent regime (the right plots of Figs. $ \ref{sx2} $ and $ \ref{sx22} $), the absence and presence of the detuning parameters show the squeezing effect in a few different intervals of time.
    \subsection{The photon statistics: Mandel parameter}\label{Mandel}

\begin{figure}[htbp]
 \centerline{\includegraphics[width=0.95\columnwidth]{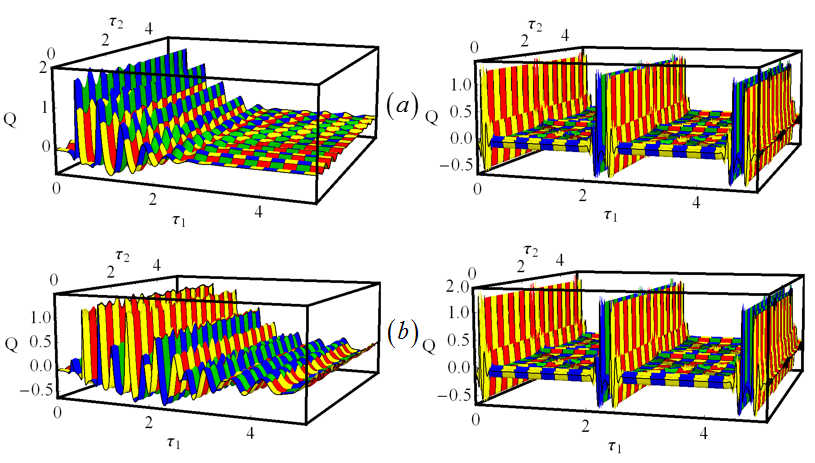}}
 \caption{Three-dimensional plots of the time evolution of the Mandel parameter in term of the scaled times $ \tau_{1}=\lambda_{2}t_{1} $ and $ \tau_{2}=\lambda_{2}t_{2} $; other parameters are chosen similar to Fig. \ref{populationinversionII}.}
 \label{mandel}
  \end{figure}

     One of the nonclassicality features in the context of quantum statistics is the sub-Poissonian behavior, the criterion which well illustrated by  the Mandel parameter in quantum optics. This quantity  is given by \cite{Mandel}:
    \begin{eqnarray}\label{mandeleq}
    Q&=&\frac{\langle(\hat{a}^{\dag}\hat{a})^{2}\rangle-\langle\hat{a}^{\dag}\hat{a}\rangle^{2}}{ \langle\hat{a}^{\dag}\hat{a}\rangle}-1.
    \end{eqnarray}
     The cases $Q>0 $ , $ Q=0 $ and $ Q<0 $ correspond to  the super-Poissonian (classical), Poissonian (for the standard coherent state) and sub-Poissonian (nonclassical) statistics, respectively. All of the required quantities in (\ref{mandeleq}) can be obtained from (\ref{number}) and (\ref{number2}).
    Accordingly, the  three-dimensional plots of the Mandel parameter versus the scaled times $ \tau_{1} $ and $ \tau_{2} $ are shown in Fig. \ref{mandel}. One can see that, this parameter takes both negative and positive values in all plots. Possessing negative values of this parameters indicates that the state vector has sub-Poissonian statistics, and so the considered interaction has transfered the classical coherent state to a nonclassical field state.
           \subsection{The Wigner quasi-probability distribution function}\label{Wigner}

\begin{figure}[htbp]
      \centerline{\includegraphics[width=0.95\columnwidth]{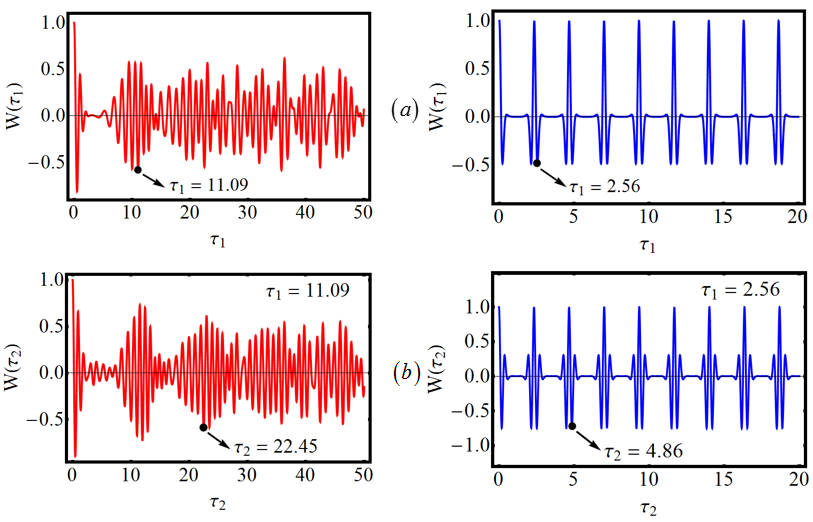}}
  \caption{The variation of the population inversion for the first in term of the scaled time $ \tau_{1}=\lambda_{2}t_{1} $ and the second atom in term of the scaled time $ \tau_{2}=\lambda_{2}t_{2} $ in the up and down rows respectively in the resonance condition for $ |\alpha|^{2}=4 $. The left plots correspond to the absence of the intensity-dependent coupling $f(n)=1$, and the right ones are plotted in the presence of the intensity-dependent coupling with nonlinearity function $f(n) = \sqrt{n}$.}
  \label{wigner}
   \end{figure}

\begin{figure}[htbp]
      \centerline{\includegraphics[width=0.95\columnwidth]{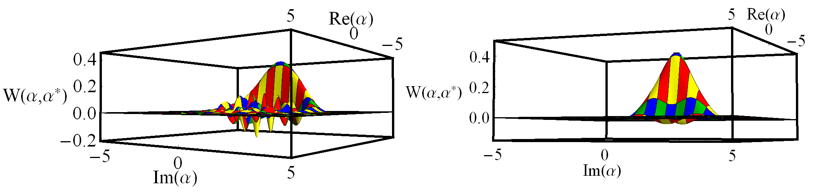}}
  \caption{The Wigner quasi-distribution function in phase space for the scaled times $  \tau_{1} $ and $  \tau_{2} $  which are defined in the Fig.  \ref{wigner}. }
  \label{wigner3D}
   \end{figure}

              This quantity is a remarkable tool to obtain enough knowledge for studying pure quantum features of a quantum system in phase space. Negativity of the Wigner distribution function in phase space is an indicator of the nonclassical feature of a specific state. This function is defined as \cite{Scully}:
          \begin{eqnarray}\label{wigner252}
          W(\alpha,\alpha^{*})&=&\frac{2}{\pi^{2}}e^{2|\alpha|^{2}}\int\langle-\gamma|\hat{\rho}_{F} |\gamma\rangle e^{-2(\gamma\alpha^{*}-\gamma^{*}\alpha)}d^{2}\gamma,
          \end{eqnarray}
           where $ |\gamma\rangle $ and $ |-\gamma\rangle $ are coherent states. Inserting the state vector of our considered system which is given by relation (\ref{wave2}) in the relation (\ref{wigner252}) and by tedious calculations, the above integral changes to the following partial differentiations for our considered system \cite{wigner}:
           \begin{eqnarray}
           W(\alpha,\alpha^{*})&=&\frac{2}{\pi}e^{2|\alpha|^{2}}\Big[\sum_{n=0}^{+\infty}\sum_{m=0}^{+\infty}\Big(\frac{(-2)^{-(n+m)}}{\sqrt{n!m!}}\nonumber\\
           &\times & \Big(A_{2}(n,t_{1},t_{2})A_{2}^{*}(m,t_{1},t_{2})+ B_{2}(n,t_{1},t_{2})B_{2}^{*}(m,t_{1},t_{2})\Big)\nonumber\\
           &\times& \frac{\partial^{n+m}}{\partial(\alpha^{*})^{n}\partial\alpha^{m}}e^{-4|\alpha|^{2}}\Big)\nonumber\\
           &+&\sum_{n=0}^{+\infty}\sum_{m=0}^{+\infty}\Big(\frac{(-2)^{-(n+m+2)}}{\sqrt{(n+1)!(m+1)!}}\nonumber\\
           &\times & C_{2}(n+1,t_{1},t_{2})C_{2}^{*}(m+1,t_{1},t_{2}) \nonumber\\
           &\times& \frac{\partial^{n+m+2}}{\partial(\alpha^{*})^{n+1}\partial\alpha^{m+1}}e^{-4|\alpha|^{2}}\Big)\Big].
           \end{eqnarray}
  In detail, the up and down plots of Fig. $ \ref{wigner} $  display the variation of the atomic inversion for the first and the second atom in term of the scaled times $ \tau_{1} $ and $ \tau_{2} $ respectively in the resonance condition for $ |\alpha|^{2}=4 $.  The left figures show the numerical results in the linear regime and the right ones deal with the nonlinear regime. In the up plots, we denote the particular the scaled times $ \tau_{1} $ in which the first atom is in its ground state, then by considering these scaled times $ \tau_{1} $ in the down plots, we show the special the scaled times $ \tau_{2} $ in which the second atom is also in its ground state, too. The Wigner quasi-probability function of our considered system for particular values of the scaled times $  \tau_{1} $ and $  \tau_{2} $, which are defined in Fig. $ \ref{wigner} $, is depicted in Fig. $ \ref{wigner3D} $. In this situation, with the constant coupling (the left plot), we see that the Wigner function gets negative values (nonclassical feature) at some finite regions in phase space; but in the presence of the intensity-dependent coupling (the right plot), one can perceive a little amount of negativity of the Wigner function in phase space.
         \section{Summary and conclusions}\label{Summary}
         In this paper, we considered the interaction between two identical V-type three-level atoms that pass from a coherent single-mode field fulfilled  a cavity consecutively using the generalized JCM with the intensity-dependent coupling between atom and field in the resonance as well as non-resonance conditions. After we obtained accurate form of the state vector of our system, the effects of the detuning parameters and the intensity-dependent coupling (by considering nonlinearity function $ f(n)=\sqrt{n} $) on the atomic population inversion, field entropy, field squeezing, photons quantum statistical and Wigner quasi-probability distribution function are examined, numerically. Summing up our  results, we achieve the following conclusions:\\
          \begin{itemize}
 \item   The intensity-dependent coupling between the atom and the field reveals the collapse-revivals  in the variation of population inversion in a clearer manner, in comparison with the constant coupling. The revivals in the intensity-dependent regime reach about its maximum value, in the resonance condition, while this is not occurred in the constant coupling. Moreover, the detuning parameters cause  a noneducable shift to higher values of population inversion in the collapse-revival pattern  in the linear regime, such that the nearly all negative values of this quantity will be disappeared.
 \item In the linear regime and in the nonresonance condition, the maximum amounts of the entropy are increased and so the entanglement between the atom and field is more visible.
 \item In the constant as well as intensity-dependent coupling regimes by considering both of the resonance and non-resonance condition, squeezing takes place appropriately in a few intervals of time at the beginnings of the interaction.
 \item  According to our numerical results, we can see that the field has the sub-Poissonian behavior either in the intensity-dependent or  constant  coupling regimes. Also, the sub-Poissonian behavior of the photons in both of the resonance and nonresonance conditions can be observed.
 \item  In order to study the quantum features of our considered atom-field system in phase space, we evaluated the Wigner quasi-probability function and find that, in the absence of detuning and  intensity-dependent regime, cavity field possess more nonclassical features according to this criteria.
 \end{itemize}
  At last, paying attention to the obtained results, we notice that either the detuning parameters or the intensity-dependent coupling have remarkable effects on the creation of nonclassical properties in appropriate situations. As an outlook of the present work, it is mentionable that the work of this paper can be accomplished for $ \Lambda $-type and $\Xi$-type three-level atoms, too. These works are performing and will be submitted in the near future.\\

         {\bf Acknowledgment} One of the authors ( E.F.) would like to thank H. R. Baghshahi for useful discussion.


%
%
\end{document}